\documentclass[12pt]{iopart}
\usepackage{graphicx} 

\def\la{\langle}
\def\ra{\rangle}
\def\beq{\begin{equation}}
\def\eeq{\end{equation}}
\def\beqa{\begin{eqnarray}}
\def\eeqa{\end{eqnarray}}
\newcommand {\fexp} [1] {\exp \left( #1 \right)}
\newcommand {\fabsq}[1] {\left| #1 \right|^2}

\newcommand {\mum}{\, \mu \mbox{m}}
\newcommand {\cms}{\, \mbox{cm/s}}

\begin{document}
\letter{Improvement by laser quenching of an ``atom diode'': a one-way barrier 
for ultra-cold atoms}

\author{A Ruschhaupt$^1$, J G Muga$^2$ and 
M G Raizen$^3$}

\address{$^1$ Institut f\"ur Mathematische Physik, TU Braunschweig, Mendelssohnstrasse 3, 38106 Braunschweig, Germany}
\address{$^2$ Departamento de Qu\'{\i}mica-F\'{\i}sica,
UPV-EHU, Apartado 644, 48080 Bilbao, Spain}
\address{$^3$ Center for Nonlinear Dynamics and Department of Physics, The University of Texas at Austin, Austin, Texas 78712-1081, USA}
\eads{$^1$\mailto{a.ruschhaupt@tu-bs.de}, $^2$\mailto{jg.muga@ehu.es}, $^3$\mailto{raizen@physics.utexas.edu}}

\begin{abstract}
Different laser devices working as ``atom diodes'' or ``one-way
barriers'' for ultra-cold atoms
have been proposed recently. They transmit ground state level atoms coming from 
one side, say from the left, but reflect them when they come from the other side.   
We combine a previous model, consisting of the stimulated Raman
adiabatic passage (STIRAP) from the ground to an excited state
and a state-selective mirror potential, 
with a localized quenching laser 
which produces spontaneous decay back to the ground
state. This avoids backwards motion, provides more control of the decay process and therefore a more 
compact and useful device.    
\end{abstract}
\nosections

Atom optics is devoted to the understanding and control of
coherent atomic waves 
interacting with electromagnetic fields or material structures, and much of its 
current development is inspired by analogies with electronics and photonics. 
Many optical elements such as lenses, mirrors, splitters, or interferometers
have been proposed and demonstrated \cite{meystre,PCGK}.
Ultra-cold atoms can also be transported
coherently along waveguides, and this opens up the possibility to develop atom
circuits or chips \cite{folman.2000, folman.2002, schneble.2003}.
To that end, basic circuit elements and operations have to be implemented; among them, an
important one is the ``diode'' or ``one-way barrier'', which lets atoms,
typically in the ground state, pass in one direction but blocks them in the
opposite direction, within a ``working'' velocity range.
Such a device may have a significant impact for trapping and cooling in
waveguides or other geometries. In a series of recent
papers \cite{rm04,raizen05,dudarev05,rm06}, the present authors and coworkers
have proposed and analyzed the properties of different laser devices and
atom-level schemes which achieve on paper the goal of one-way
transmission. The possibility to use them for phase-space compression in a
cooling procedure complementary to the existing ones \cite{raizen05,dudarev05}
is an exciting prospect that deserves further research and motivates several
ongoing experimental and theoretical efforts in which 
this paper may be framed.

In this letter, we tackle a significant practical
improvement for some of the proposed diode implementations: the use of a
quenching laser to induce at will excited state decay. 
A good diode must involve an irreversible decay step so that time-reversed
trajectories associated with backwards motion in the ``forbidden'' direction
do not occur. For example, in a simple one dimensional (1D) diode scheme,
ground state  atoms coming from the left within a broad
velocity range are transmitted to the right in some excited state whereas 
ground state atoms from the right are blocked by a state-selective mirror. 
While this may be enough for some purposes, the absence of an irreversible
decay 
from the excited state would mean, according to the unitarity of the collision
matrix,
that an excited atom could cross the device leftwards.  
Since excited atoms could come back towards the diode because of collisions or
external forces,
a truly one-way device, reliable when atom returns are possible,  requires the irreversible decay of the 
excited atoms in a time scale smaller than the return time. 
This condition may occur naturally by a direct spontaneous decay.
Nevertheless, long lived excited states need a long time to decay or,
equivalently, the atom has to travell a long distance to decay, 
which leads to a spatially wide ``atom diode'', a  
drawback for cooling applications. 
To remedy this problem, it is possible 
to force at will the irreversible decay via a quenching laser, as it  
is explored and demonstrated here.

\begin{figure}
\begin{center}
\includegraphics[width=0.35\linewidth]{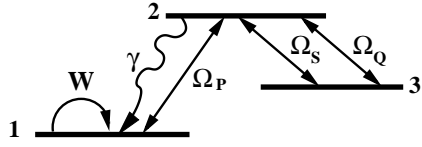}
\end{center}
\caption{\label{fig1} Atomic level structure for the diode model.}
\end{figure}

\begin{figure}
\begin{center}
\includegraphics[width=0.65\linewidth]{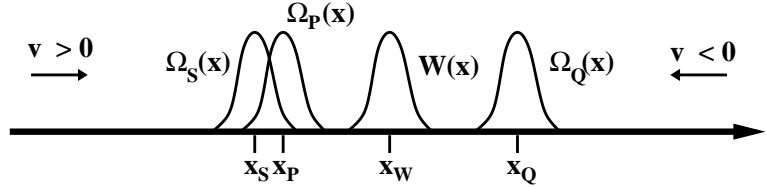}
\end{center}
\caption{\label{fig2} Spatial localization of the lasers. 
In the calculations $x_S = -15\mum$, $x_P=15\mum$, $x_W=85 \mum$, and
$x_Q=155\mum$.}
\end{figure}

In our model the atom has the three-level structure 
of Fig. \ref{fig1} with spontaneous decay from $2\to 1$. 
%
We assume that the atomic motion is effectively 1D and use a combination of
``Stokes'' and ``pumping'' STIRAP lasers \cite{STIRAP} with position dependent Rabi
frequencies $\Omega_S$, and $\Omega_P$, a state-selective
mirror laser, and a quenching laser with Rabi frequencies $W$, and
$\Omega_Q$, respectively. Their spatial location  is shown in Fig. \ref{fig2}.
%
A similar model of the three-level atom diode has been proposed in \cite{rm04}
without quenching potential, but with spontaneous decay from $3\to1$.
In that case the decay rate $3\to 1$ must be small enough such that the
probability for spontaneous decay is negligible
before the rightwards-traveling
atom has passed the state-selective mirror (see \cite{rm04} for details), but 
the small decay rate implies a spatially wide ``atom diode'' region.

For a full description of the atomic dynamics corresponding to 
Figs. \ref{fig1} and \ref{fig2}, including the possibility
of  several spontaneous emission cycles, the following 1D master equation 
must be examined \cite{qj,Hensinger01,ROS}:
\begin{eqnarray}
\frac{\partial}{\partial t} \rho
&=& - \frac{i}{\hbar} [H_{3L}, \rho]_-
- \frac{\gamma}{2} \{|2\ra \la 2|,\rho\}_+
\\
&+&\!\gamma\!\int_{-1}^{1} \!\!du \frac{3}{8} (1+u^2) 
\fexp{i\frac{mv_{rec}}{\hbar} u x}
\,|1\!\ra 
\la 2|\rho|2\ra  \la 1|\,
\fexp{-i\frac{mv_{rec}}{\hbar} u x}, \nonumber
\label{master_eq}
\end{eqnarray}
where $v_{rec}$ is the recoil velocity, $m$ is the mass, 
and the initial condition is taken as a pure state.
Using $|1\rangle \equiv (1\,0\,0)^T$, $|2\rangle \equiv (0\,1\,0)^T$,
and $|3\rangle \equiv (0\,0\,1)^T$, the non-hermitian
conditional Hamiltonian is
\begin{eqnarray}
H_{3L} = \frac{p_x^2}{2m} + \frac{\hbar}{2} \left(\begin{array}{ccc}
W(x) & \Omega_P(x) & 0\\
\Omega_P (x) & -i \gamma & \Omega_S (x) + \Omega_Q (x)\\
0 & \Omega_S (x) + \Omega_Q(x) & 0
\end{array}\right).
\label{ham}
\end{eqnarray}
All potential terms are chosen as Gaussian 
functions with equal widths $\Delta x = 15 \mum$ and maximal heights
$\hat{\Omega}_P$,
$\hat{\Omega}_S$, $\hat{W}$, $\hat{\Omega}_Q$, 
\begin{eqnarray*}
&W (x) = \hat{W} \;\Pi(x,x_W),\quad \Omega_S(x) = \hat{\Omega}_S \;\Pi(x,x_S),&\\
&\Omega_P (x) = \hat{\Omega}_P\; \Pi (x,x_P),\quad \Omega_Q (x) = \hat{\Omega}_Q\; \Pi (x,x_Q)&,
\end{eqnarray*}
where $\Pi (x,x_0)=\exp[-(x-x_0)^2/(2\sigma^2)]$. 
It is important that the distance $x_W - x_P$ is not too small
because otherwise the mirror
potential spoils the STIRAP transfer, 
and of course $x_Q \gg x_W$ must hold such that the atom crossing rightwards has passed the mirror potential before
it decays to state $1$.
 
In the quantum-jump approach, the master equation (\ref{master_eq}) 
is solved by averaging over 
``trajectories'' with time intervals in which the wave function evolves with the
conditional Hamiltonian
interrupted by random jumps (decay events) \cite{qj}.
Therefore the dynamics before the first spontaneous photon emission is
described by a simple Schr\"odinger equation using
the conditional Hamiltonian $H_{3L}$.
%
%
Note that the STIRAP process $1\to3$ is characterized by its
stability with respect to atom velocity and the fact that
it is not  much affected by the decay $2\to1$, since the adiabatic state 
realizing the transfer is
not composed by 2 at all. Also, because of the position of the quenching laser
promoting  spontaneous decay from 2 and the assumed level structure,
it will be extremely unlikely to have more than one pumping-emission cycle.
Thus, the dynamics under the conditional Hamiltonian will be essentially enough 
to determine the quality and behaviour of the diode. In particular, 
it will help us to select good values for the laser parameters. 
So let us examine the stationary Schr\"odinger equation with $H_{3L}$. 
Using $\alpha$ and $\beta$ to denote the channels associated with the internal states, $\alpha=1,2,3$,
$\beta=1,2,3$, let us denote as 
%
$R_{\beta\alpha}(v)$ the reflection amplitude for an atom incident 
with velocity $v$ (positive or negative, see Fig. \ref{fig2}) in channel  
$\alpha$ and reflected in channel $\beta$. 
Similarly, let $T_{\beta\alpha}(v)$ be the transmission amplitude 
from $\alpha$ to $\beta$. 

First we examine the case without quenching potential ($\Omega_Q = 0$)
but with $\gamma > 0$. The parameters $\hat{\Omega}_P=\hat{\Omega}_S$
have to be chosen large enough such that all ground state atoms incident from
the left are transmitted into state 3 (i.e. $\fabsq{T_{31}(v)}\approx 1$ for
$v>0$) whereas from the right they are reflected in a
velocity interval (i.e. $\fabsq{R_{11}(v)}\approx 1$ for
$v<0$).
%
\begin{figure}
\begin{center}
\includegraphics[angle=0,width=0.5\linewidth]{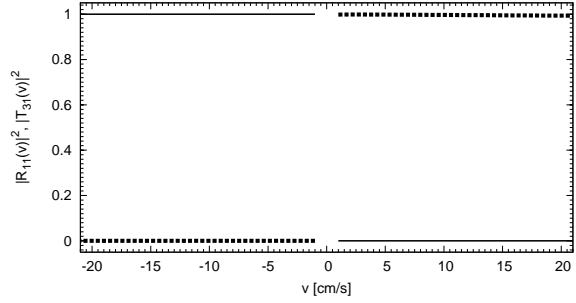}
\end{center}
\caption{\label{fig3}Reflection probability $\fabsq{R_{11} (w)}$ (solid line),
transmission probability $\fabsq{T_{31} (v)}$ (dashed line). Here and in the 
following figures, $\hat{\Omega}_P = \hat{\Omega}_S = 10^6/s$,
$\hat{W}=2\times 10^7/s$, $\hat{\Omega}_Q = 0$, $\gamma = 10^5/s$, 
and $m$ is the mass of Neon.}
\end{figure}
%
%
Figure \ref{fig3} shows that this is true for the chosen parameters. 
Notice the stability of the results with
respect to velocity because of the adiabatic passage.

Without a quenching laser, the process 3$\to$1 from the right, however, would 
be also permitted and in fact very efficient. We shall now include the
quenching laser and calculate again the transmission probability for the
process $1\to 3$, namely, $|T_{31}(v)|^2$ for $v>0$.   
%
\begin{figure}[t]
\begin{center}
\includegraphics[width=0.5\linewidth]{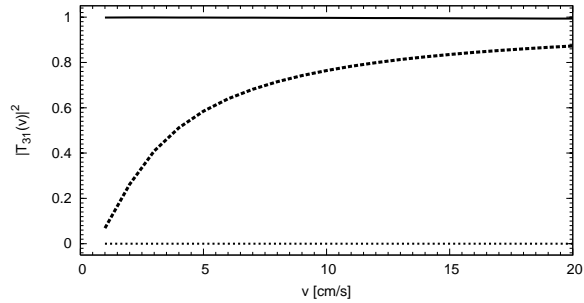}
\end{center}
\caption{\label{fig4} Transmission probability $\fabsq{T_{31}(v)}$
for different $\hat{\Omega}_Q$;
$\hat{\Omega}_Q = 0$ (solid line),
$\hat{\Omega}_Q = 10^{4}/s$ (thick dotted line),
$\hat{\Omega}_Q = 10^{5}/s$ (dotted line).} 
\end{figure}
%
Figure \ref{fig4} shows that by increasing $\hat{\Omega}_Q$ it is possible to 
suppress totally $|T_{31}(v)|^2$. Since reflection for ground state 
atoms coming from the left is still negligible, 
this means that the norm of a wave function of atoms incident in level 1 from the
left is absorbed totally under the time-dependent Schr\"odinger equation
with the non-hermitian Hamiltonian $H_{3L}$.
Physically, nearly
all atoms incident in state 1 from the left will emit a spontaneous photon.
Because the emission will take place very likely close to the
quenching potential, the atom will finally
travel to the right in the ground state, as confirmed below 
by solving the master equation. 
All other probabilities represented in Fig. \ref{fig3} remain unchanged; in particular, the atoms coming from the right in the ground state 1 are  
unaffected by the quenching laser but blocked by the selective mirror
potential. Notice also that, even atoms incident from the right in 3 
will be blocked if
the decay provoked by the quenching laser occurs in a time scale which is
enough for the atom to arrive at the mirror potential in state 1.

In principle, if $\hat{\Omega}_Q$ is too large there could be reflection
of an atom  incident in level 1 from the left
but this possibility plays no role for the parameters examined here.
So the quenching laser has a range of intensities and velocities for
which the desired effect (perfect decay of atoms in level 3) is stable.
This regime is also discussed for a single pumping laser without diode
in \cite{damborenea.2003}. 

To confirm that the quenching laser is indeed improving the device  
significantly we shall solve the 1D master equation (\ref{master_eq}) 
using the quantum trajectory approach.
%
%
\begin{figure}[t]
\begin{center}
\includegraphics[angle=0,width=0.49\linewidth]{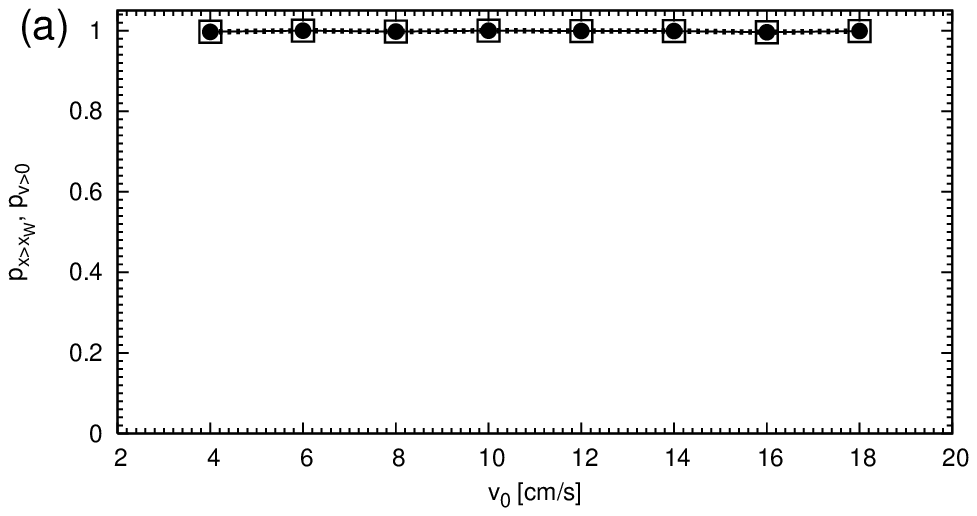}
\includegraphics[angle=0,width=0.49\linewidth]{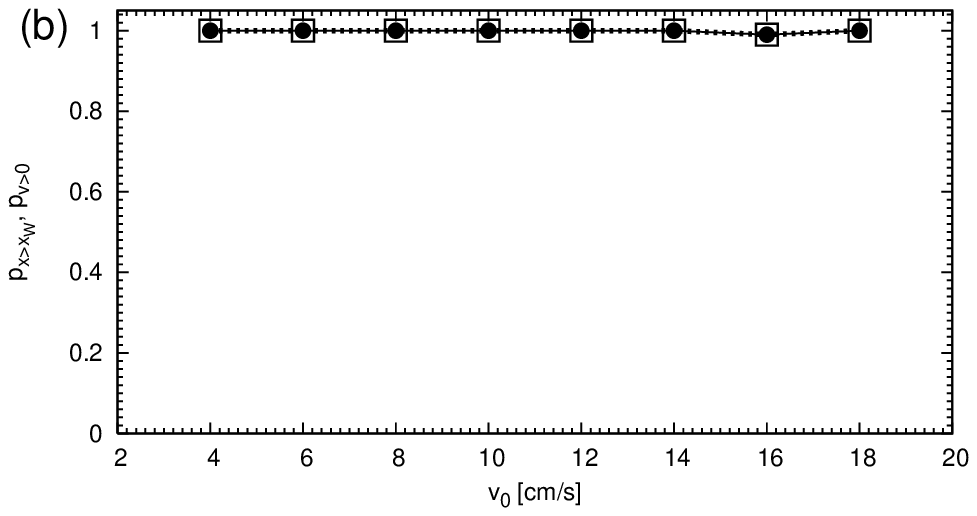}
\end{center}
\caption{\label{fig5} Probabilities $p_{x>x_W}$ (filled circles)
and $p_{v>0}$ (boxes) versus $v_0$.  $\Delta v_0 = 0.1 \cms$, 
$x_0 = -115 \mum$ ($v_0>0$) or $x_0 = 255 \mum$ ($v_0<0$).  $t_{max} = 600 \mum / v_0$, $\hat{\Omega}_Q = 10^5/s$.
(a) $v_{rec} = 0$ (averaged over $N=1000$ trajectories),
(b) $v_{rec} = 3 \cms$ (averaged over $N=100$ trajectories). 
}
\end{figure}
%
The initial state at $t=0$ is $\rho(0)=|\Psi_0\!><\!\Psi_0|$,
namely a Gaussian wave packet with mean velocity $v_0$ in the ground state, 
\begin{eqnarray*}
\Psi_0 (x) = \frac{1}{N} \left(\begin{array}{c} 1\\0\\0\end{array}\right)
\exp\left[-\frac{\Delta v_0 m}{2\hbar} (x-x_0)^2 
+ i \frac{v_0m}{\hbar} x\right],
\end{eqnarray*}
where $N$ is a normalization constant. 
Let $t_{max}$  be a large time such that the resulting wave
packet 
$\Psi_j (t_{max})$ of almost every quantum ``trajectory'' $j$ separates
into  right and left moving parts far from the interaction region.
By averaging over all trajectories we get 
%
$
p_{x>x_W}\!\!=\!\!\int_{x_W}^\infty  dx\, \la x|\rho_{11}(t_{max})|x\ra,
$
%
which is the probability to find the atom on the right-hand side and in the
ground state, and also  
%
$
p_{v>0}\!\!=\!\!\int_{0}^\infty  dv\, \la v|\rho_{11}(t_{max})|v\ra,
$
%
which is the probability to find the atom moving to the right at time
$t_{max}$ in the ground state.
The results, shown in Fig. \ref{fig5}, confirm the prediction of
the stationary Schr\"odinger equation with the
conditional Hamiltonian: a perfect transfer is produced from ground to ground
state.
The error bars (defined by the difference between
averaging over $N$ and over $N/2$ trajectories) are smaller than the
symbol size.

In summary, a quenching laser has been included in an atom diode laser device.
Using quenching to force decay to the ground state after crossing the diode
provides a significant improvement over previous schemes based on small direct
spontaneous decay. With quenching, possible crossings in the wrong direction are suppressed and the atom which has crossed the diode in the allowed direction needs less time to decay, or in other words, the spatial region occupied by the atom diode is much narrower than in the case of direct decay. Quenching could be used in realistic atomic systems such as calcium, strontium, or ytterbium which are currently of interest for frequency standards.

\ack{JGM acknowledges support from 
``Ministerio de Ciencia y Tecnolog\'\i a-FEDER''
(BFM2003-01003), and
UPV-EHU (Grant 15968/2004). MGR acknowledges support from NSF, the R. A. Welch Foundation, and the S. W. Richardson Foundation and the US Office of Naval Research, Quantum Optics Initiative, Grant N0014-04-1-0336.}

%

\end{document}